# Community paper "Radio astronomy infrastructures"

Eduardo Ros[1], Dominik J. Schwarz[2], & Christian Vocks[3]
on behalf of the German astronomical community

1: Max-Planck-Institut für Radioastronomie, Bonn; 2: Universität Bielefeld; 3: Leibniz-Institut für Astrophysik Potsdam

## Executive summary

Radio astronomy has experienced phenomenal progress in recent years due to advances in digital technologies and processing speed, the development of new technologies, and the prospect for new powerful facilities  A new generation of radio interferometers is opening new windows to the Universe.  LOFAR, the world largest telescope, extended the classical radio window at low frequencies.  ALMA started a new era at millimetre wavelengths and the planned SKA will revolutionize the sciences of the Universe, well beyond the traditional limits of astronomy.  Radio astronomy research is making leaps with enhancements in resolution, sensitivity, and image fidelity.  In Germany, the community has access to excellent facilities and training opportunities.  The Effelsberg telescope remains the flagship of radio astronomical research at centimeter wavelengths and serves as a test bed for new technologies, LOFAR has notably expanded the German community and now includes six German stations and a LOFAR long term archive coordinated by the GLOW consortium.  The SKA will be a transformational astronomical facility in the coming decade(s), and the German community is looking forward to broadly participate in SKA-enabled research.

## 1. Introduction

### Science and technique

Radio astronomical techniques allow us to observe the sky at the largest accessible wavelength (or smallest accessible frequencies) of electromagnetic radiation (light).  Radio astronomy was born in the 1930s and gave rise to several ground-breaking discoveries since, e.g. discovering the existence of gravitational waves indirectly from the double pulsar (Hulse & Taylor, Nobel Prize 1993) or establishing the isotropy of the Universe and the discovery of the cosmic microwave background (Penzias & Wilson, Nobel Prize 1978; Mather & Smoot Nobel Prize 2006).  Radio astronomy also led to important technological spin off, such as the WLAN protocol.

A major advantage of radio astronomical techniques is that observations are possible from the Earth's surface. Unlike optical telescopes, the technique works during daylight time and even with cloudy skies.  The observing window ranges from wavelengths of several meters, where the main limitation comes from Earth's ionosphere reflecting waves with frequencies less than its plasma frequency (around 10 MHz), to the sub-millimetre range (around frequencies of 100 GHz), where water vapour in the atmosphere affects observations, forcing telescopes to be installed at high, dry sites, or to use balloons or space borne probes for detailed studies of the microwave sky.  On ground, observational techniques can use extremely large collecting surfaces, but suffer in principle from low resolution $\theta$, given the long wavelengths $\lambda$ used with respect to the aperture dimensions $D$ ($\theta \approx \lambda/D$).  This problem is overcome with the use of interferometric techniques, which make use of the aperture synthesis principle (Ryle & Hewish, Nobel Prize 1974), both in connected facilities such as the Jansky Very Large Array (JVLA) in New Mexico, USA, or the Low Frequency Array (LOFAR) in Central Europe, or by using independent elements synchronised with atomic standards, applying very-long-baseline interferometry (VLBI).

Radio astronomy uses various observational techniques.  Basically, any radio telescope system hosts a frontend or receiver and a backend system, where the signal is electronically processed and analysed.  The frontend can be a single receiver (one pixel or a single antenna) or a focal plane array to record simultaneously different data points (like a CCD in optical imaging), and can register all polarisation properties with linear or circular feeds.  The backend can split the signal in the time domain, for instance, to analyse rapidly changing, periodic signals from neutron stars, and in the frequency domain, to analyse spectral lines in emission and absorption.  Images from a single telescope can be obtained by overlapping multiple scans pieced together in a mosaic.  Using radio telescopes in interferometric mode yields detailed images after Fourier deconvolution, surpassing any other astronomical technique in resolution, reaching values of $10^{-10}$ rad (or 10 µas) with mm- or space-VLBI (e.g. by combining data from ground based observatories with the Russian radio space mission RadioAstron).  Basically, a telescope uses four observing modes: continuum (which can include on-offs, cross scans, polarisation, and mosaic imaging), time domain (including the study of radio pulsars and the recently discovered fast radio bursts), spectroscopy (atomic fine structure and molecular lines can be observed in the radio), and VLBI.

Here we focus on the facilities operating above 1 cm wavelength, and mention briefly telescopes operating at shorter wavelengths, which will be addressed in a parallel contribution. Facilities for observation of the cosmic microwave background will be addressed in the contribution on space borne probes and in the contribution on infrared and sub-millimetre facilities.

### Strategic considerations

AstroNet, the European astronomy network established in the 2000s, produced a Science Vision[i] in 2007, which was updated in 2012. This project established an Infrastructure Roadmap in 2008, updated in 2014, naming the Extended European Large Telescope (E-ELT) of the European Southern Observatory (ESO) and the Square Kilometre Array (SKA) as the two top priorities for European astronomy.  This is also acknowledged by the European Strategy Forum on Research Infrastructure (ESFRI), which lists E-ELT and SKA as its two astronomy landmarks in its 2016 roadmap.

AstroNet established a panel on radio astronomy, the European Radio Telescope Review Committee[ii] (ERTRC), which delivered its report in 2015.  In this report, radio astronomy is considered as a healthy and blossoming branch of science in transition to large, and more complex facilities.  Challenges such as big data, accessibility and managing a growing community are identified.  The ERTRC recommends pursuing research in the areas of neutral atomic hydrogen at high redshift, strong gravity, compact structures, cosmic magnetism, and molecular astrophysics.  Several existing facilities are recognized as essential and complementary to the SKA.  The European VLBI Network (EVN), and its managing institute, the Joint Institute for VLBI in Europe (JIVE) were highlighted, as well as the International LOFAR Telescope (ILT).  Additionally, the science case for single dishes was stressed, namely, surveys, monitoring or transient search.

### German radio facilities
The German landscape has the following radio astronomy landmarks:
- The **Effelsberg 100-m radio telescope** operated by the Max-Planck-Institut für Radioastronomie, built in the 1970s and continuously improved since then (see Figure 1). In recent years, new adaptive optics and receiver upgrades have been installed. The new 1.3 cm (K-band) and 3.2-7.5 cm (broad C-band) systems are ready, and two new receivers in the 6-9mm (Q) and 1.7-2.5 cm ($K_u$) ranges will be installed in 2016. The new systems provide broad frequency coverage, all with ultra-wide bandwidths.  A focal plane array is going to be deployed at 18/21 cm.  Digital techniques are being vastly improved. Optical fibre allows high data bit rates for VLBI to other antennas.  The wavelength coverage ranges from 3 mm to 90 cm. Owing to its large collecting surface, Effelsberg is a key element in interferometric arrays. Apart from enabling cutting-edge research, the facility is also an incomparable testbed for new technologies, given its large collecting surface, where a single element can be tested in comparison with a multi-element interferometer.

- **Six international LOFAR stations** are now installed, in Effelsberg, Jülich, Norderstedt, Potsdam, Tautenburg, and Unterweilenbach (see Figure 3), coordinated by the German Low Wavelength Consortium (GLOW)[iii]. A **LOFAR Long Term Archive** (LTA) is operated by the FZ Jülich, which grows currently at a rate of 3 PB/yr. The International LOFAR telescope (ILT) has been active since 2013 and offers wavelength coverage from 1.3 m to 30 m. Germany is the second biggest contributor to the ILT, which is todays largest telescope in the world (both in terms of collecting area and in terms of data rates).
- An **ALMA Regional Centre (ARC)** is operated by the Universities Bonn and Cologne. ALMA is the Atacama Large Millimeter/submillimeter Array operated by ESO in Chile. Details are provided in the section on infrared and submillimetre facilities.
- Contribution to wider **networks**, namely: Effelsberg participates in the **EVN (European VLBI Network)**, the High-Sensitivity Array (HSA), the Very Long Baseline Array (VLBA), the Global VLBI Network, the Global mm-VLBI Array (**GMVA**), and the European/International Pulsar Timing Arrays (**EPTA/IPTA**); the GLOW institutes participate at the International **LOFAR** Telescopes; the Argelander-Institut für Astronomie participates in the Dutch **APERTIF** project
- **Geodetic** infrastructure using radio astronomy techniques, like the telescopes in Wettzell (Bavarian Forest) and O'Higgins (Antartica).
- **Atmospheric and ionospheric** research facilities such Kühlungsborn.

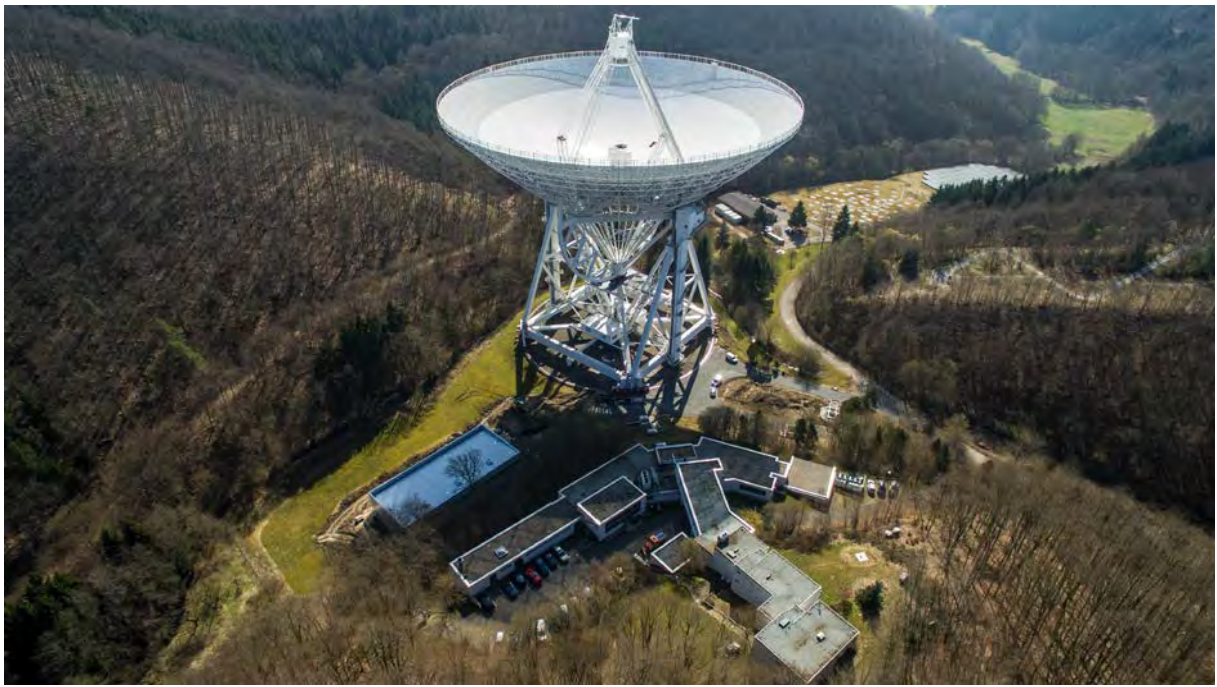

**Figure 1: The 100-m radio telescope Effelsberg near Bad Münstereifel. On the top, right, the DE601 international LOFAR station (Credit: MPIfR)**

Germany also participates in other radio astronomy facilities. The MPIfR has a bilateral agreement with JIVE (Joint Institute for VLBI in Europe), which was established in December 2014 as a European Research Infrastructure Consortium (ERIC). The MPIfR also contributes with a special agreement to the VLBA (to become the Long Baseline Observatory in late 2016).  A cooperation agreement exists with the space VLBI mission RadioAstron (operated by RosCosmos).  The MPIfR has committed to collaborate with the SKA pathfinder MeerKAT in South Africa by providing receivers at 13 cm wavelength.  In the millimetre regime, collaboration exists with the Institut de Radioastronomie Millimétrique (IRAM), with a 40% of MPG funding and telescopes NOEMA in France and Pico Veleta in Spain, the MPIfR holds 51% of the APEX telescope in Chile, and through the German share in ESO, participates at the ALMA observatory in Chile.  Until June 2015, Germany was a member of the SKA organisation (see below).

This impressive range of facilities accessible to German researchers reflects the strength and leading role of the German community.

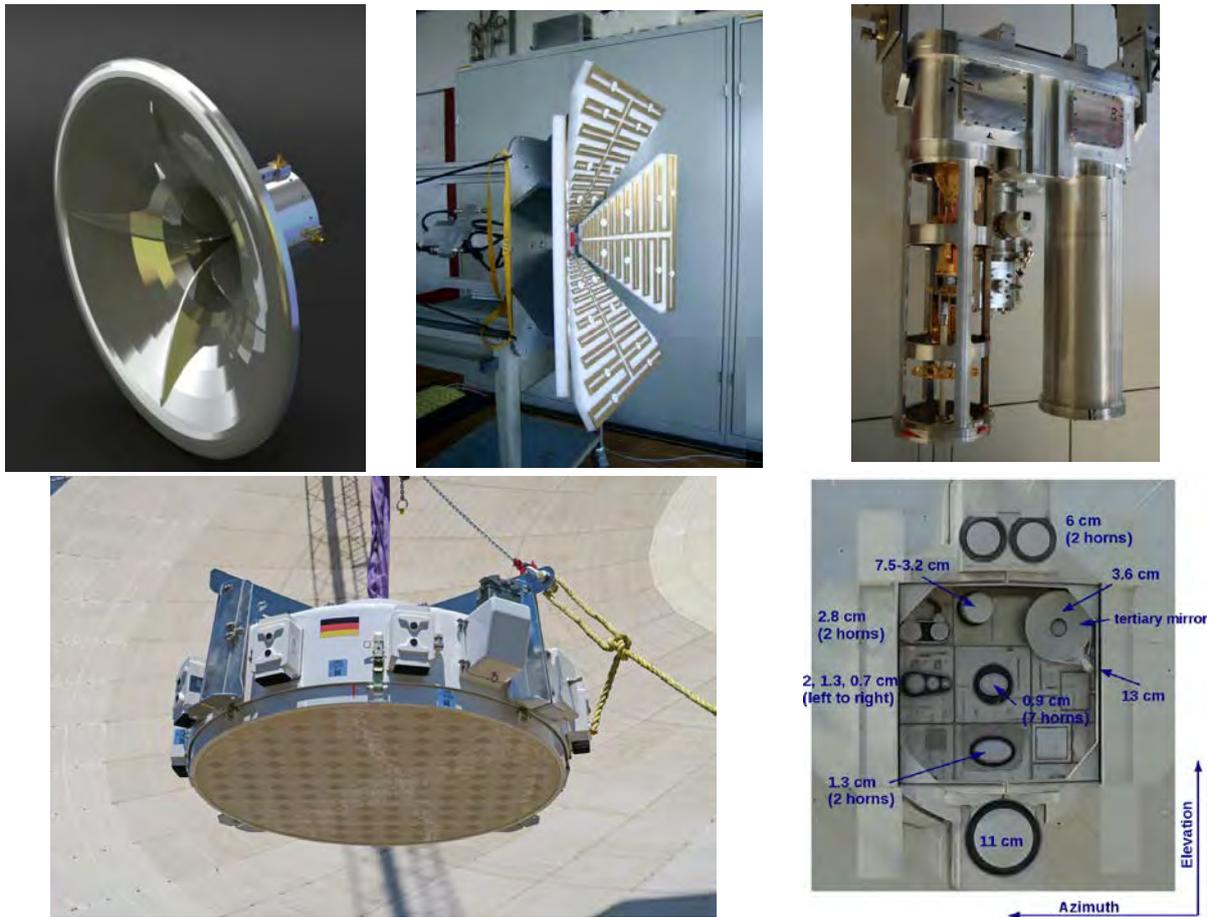

**Figure 2: Examples of recent technical developments in Effelsberg.  Top left: new ultra-broad band receiver (ll from 10cm to 50cm); Top centre: 50cm broadband receiver.  Top right: 1.3 cm cryostat. Bottom left: New 21cm Focal Plane Array.  Bottom right: suite of receivers in the secondary focus of the telescope.**

## 2. Upcoming Facilities in the Coming Decade

The world's major facility will be the Square Kilometre Array. The SKA will be built in phases and SKA Early Science will start in 2020. Complementary upgrades to the current world leading interferometers JVLA and LOFAR (LOFAR 2.0) are planned.

The world's biggest single dish will be the FAST, which is currently being built in China. Several smaller research infrastructures focus on the discovery and mapping of the Epoch of Reionization and Cosmic Dawn, such as Paper or MWA and on HI intensity mapping at lower redshifts, such as BINGO and CHIME. Multi-purpose radio interferometers are the SKA pathfinders MeerKAT and ASKAP.

The Event Horizon Telescope is a global collaboration including Germany teaming up radio telescopes operating at wavelengths of 0.8–1.3 mm in an ad-hoc VLBI experiment with the goal of observing directly the immediate environment of a black hole with angular resolution comparable to the event horizon.  Included telescopes are Pico Veleta, Plateau de Bure, APEX and beamformed ALMA, among others.  The two main astrophysical targets are Sgr A* and M 87.

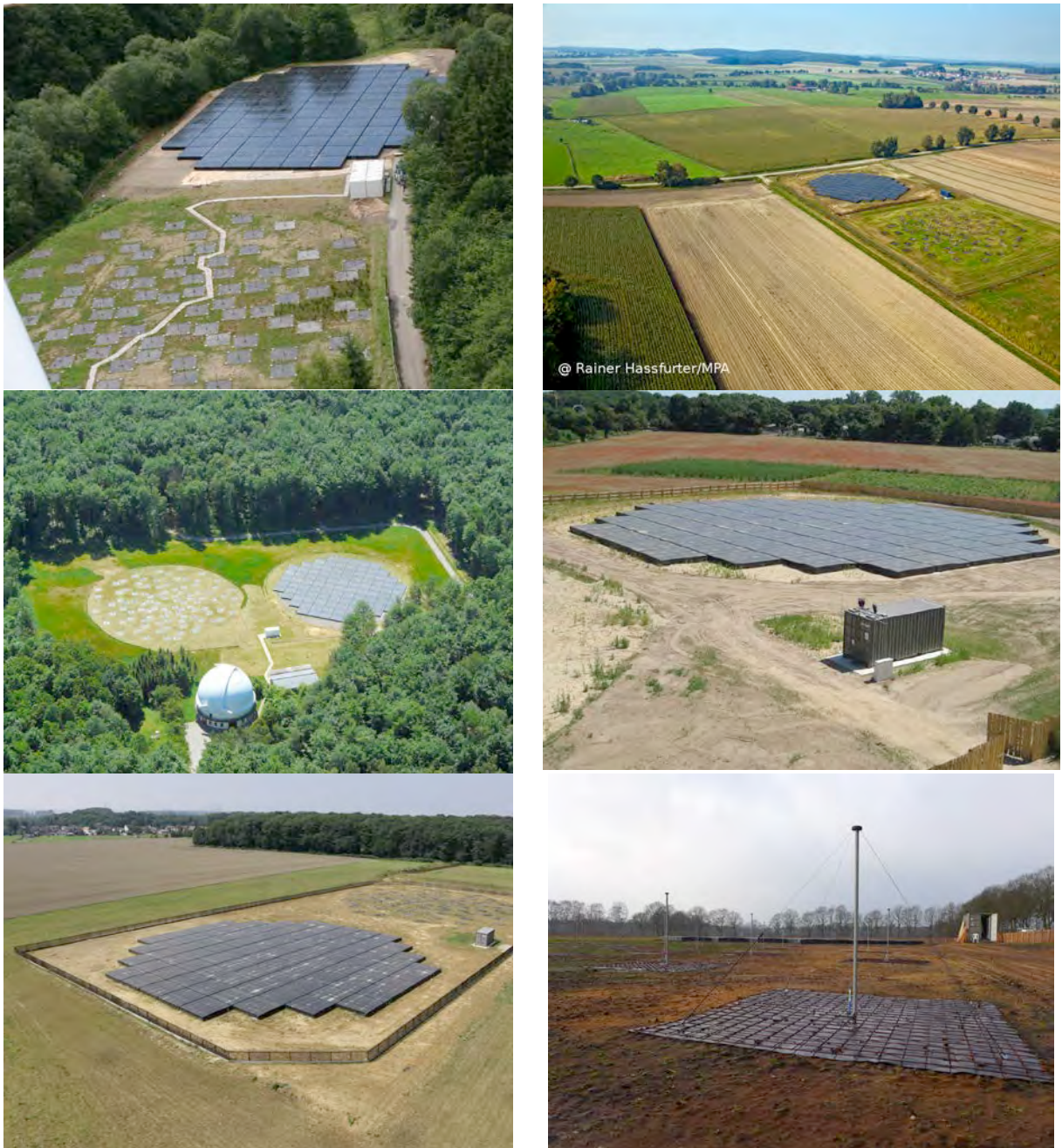

**Figure 3 - German LOFAR stations.  From the left to right, top to bottom: Effelsberg (MPI für Radioastronomie), Unterweilenbach/Garching (MPI für Astrophysik), Tautenburg (Thüringer Landessternwarte), Potsdam-Bornim (Leibniz-Institut für Astrophysik Potsdam), Jülich (Ruhr-Universität Bochum and Forschungszentrum Jülich), and Norderstedt (Universität Bielefeld and Universität Hamburg).**

From the perspective of the German astrophysical community, it will be essential to participate in the most important international project, the SKA. Besides participating in SKA it will be essential to continue the strong German LOFAR involvement in order to harvest the fruits of a decade of technology and infrastructure development, to train a new generation of researchers and engineers, as well as to further develop LOFAR to complement the science potential of SKA. In the coming decade, Effelsberg will continue with upgrades, for instance with a new focal plane array at 18/21 cm wavelength.

### Effelsberg
In the coming decade, the Effelsberg 100-m radio telescope operated by the MPG will continue to be a world-class instrument. . Upgrades, like the mentioned deployment with a focal plane array at 18/21cm wavelength, are ongoing or planned. Figure 2 shows recent technical developments and a picture of the suite of receivers in the secondary focus.

### LOFAR 2.0
The plans of upgrading LOFAR are established in the LOFAR 2.0 project. The goal is to remain unique and competitive, and the implementation of the upgrade should be carried out between 2018 and 2023. The major science cases for this upgrade are cosmic magnetism, cosmic dawn, and solar physics (see Section 5. Dominant Science Cases). The upgrade will allow for simultaneous low-band and high-band array observations in all stations, replace present low-band antennas, to cope with new radio-frequency interference and to increase the low band array (3.7 m to 30 m wavelength) sensitivity by an order of magnitude, and increase the filling factor in the core station.

### SKA
The SKA is a global research infrastructure with two sites in Southern Africa and Western Australia. Two interferometers will be built, SKA-LOW with wavelengths of 85 cm to 6 m in Australia, and SKA-MID in the range of 21 cm to 85 cm in Africa. The headquarters of the SKA organisation are located in Jodrell Bank in the United Kingdom. The phase 1 of the SKA project has a cost cap of 650 M€, and the annual running costs are expected to be in the range of 10−12 % of the total. Figure 4 illustrates the SKA in comparison with the world's most sensitive radio telescopes. The timeline for the SKA included science priorisation in 2014 and a rebaselining in 2015. Negotiations to found an intergovernmental organisation started in 2015, including Italy, Australia, South Africa, the United Kingdom, Canada, China, India, New Zealand, The Netherlands, and Sweden. A critical design review will be performed in 2016 and the intergovernmental SKA organisation will be established in 2017, as the tender and procurement will start. The construction is planned for 2018−2024, with early science expected around 2020.

Germany, after exiting the SKA Organisation in 2015, is at present invited as a guest with observer status. The German community continues to be involved in SKA, participating very actively in science working group activities, in design consortia work and at advisors to the project. The German ministry of education and research called for a proposal for a National Roadmap Process, due in January 2016. A consortium of German institutions has submitted a proposal for a German contribution to the SKA. It is planned that a German SKA Consortium will be established in 2016 coordinating both scientific and industrial activities. The consortium will be based on the current GLOW SKA working group, so that GLOW will continue to play a major role. In this framework, and following the fair-return principle, having Germany joining the SKA again, the German industry partners should be able to take part in procurement and tender by 2018.

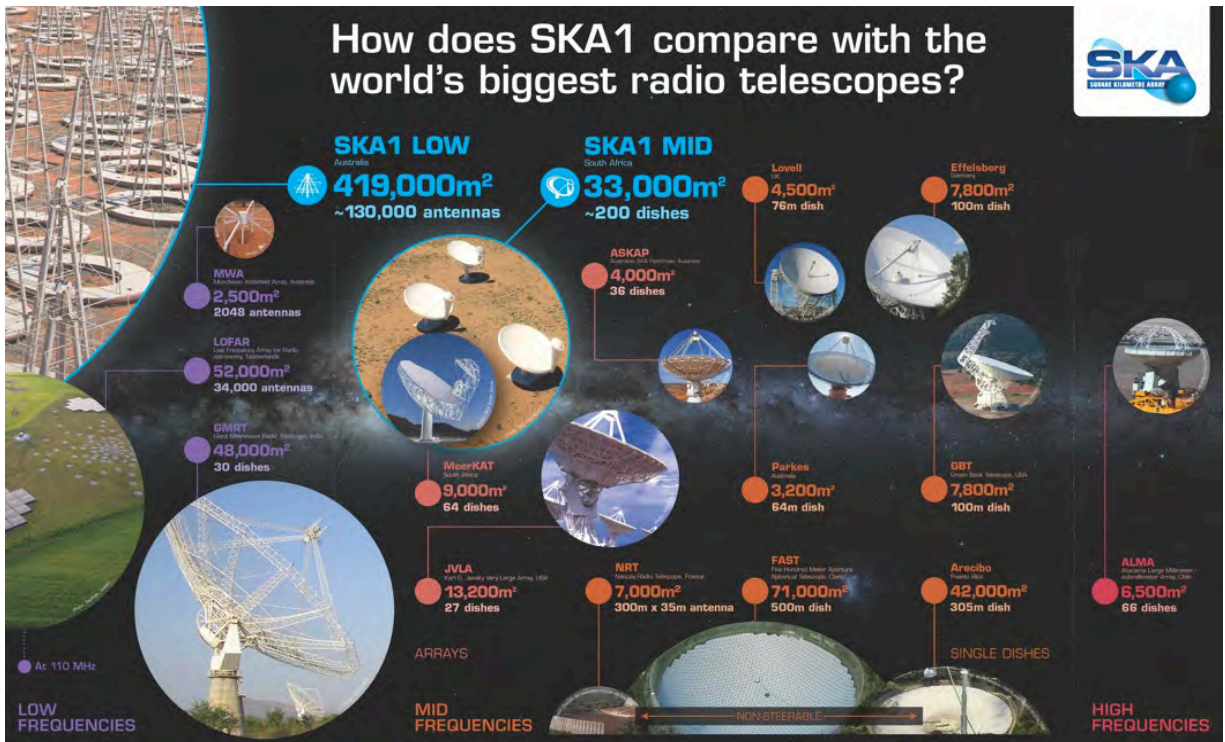

**Figure 4: Comparison in sensitivity of the SKA1 with the largest radio telescopes world wide (credit: SKAO, http://www.skatelescope.org)**

## 3. Main Achievements in the Past Decade

German radio astronomy is a growing community. After a traditional focus at the MPIfR in Bonn and Bochum, the deployment of LOFAR, the GLOW consortium, ALMA and several new working groups in different locations (e.g., at the MPI für extraterrestrische Physik and at the Universities of Bielefeld, Hamburg or Würzburg) have broadened the German radio astronomical community and its geographical balance.

A review of the big scientific achievements in the last ten years in the German radio astronomy landscape above 1 cm wavelengths shows several highlights (and excluding the German contributions to the study of the cosmic microwave background which is to be covered in a parallel document of this memorandum). These include:

- Discovery of the most massive neutron star with Effelsberg and Arecibo[iv]
- Measurement of trigonometric parallaxes of star forming regions in the framework of the BeSSeL project, using the VLBA[v]
- The HI Nearby Galaxy Survey (THINGS)[vi]
- The MOJAVE monitoring program with blazar kinematics, showing the connection of radio and γ-ray emission and the role of magnetic fields in AGN from VLBA monitoring over the last 20 years[vii]
- Discovery of a magnetar near the Galactic Center[viii]
- The monitoring results of blazars from the F-GAMMA program in Effelsberg[ix]
- The Effelsberg-Bonn HI (neutral atomic hydrogen) survey[x]
- Pulsar timing using combined large telescopes[xi]
- Magnetism studies with Effelsberg, e.g., imaging of NGC 6946[xii]
- First redshift measured for a fast radio burst[xiii]
- Composition of ultra-high-energy cosmic rays (LOFAR)[xiv]
- Cosmic radio dipole from NVSS and WENSS[xv]
- Mapping of magnetic fields around spiral galaxies (M81)[xvi]
- Observations of the radio lobes of Virgo A at 30 MHz[xvii]
- Spectroscopic Solar imaging by the Potsdam-led LOFAR Solar Key Science Project[xviii]

Other programs have a large impact, including studies with Effelsberg of polarisation sources, cosmic microwave background (contributions to the data analysis of Planck and the South Pole Telescope), ammonia from infrared dark clouds, magnetic field strength in galaxies, pulsars and transients, etc.

The recipe for success were the technical developments in Effelsberg, the involvement in international collaborations, and the efforts in new areas of radio astronomy such as cosmological studies, multi-band and multi-messenger approaches, etc. A key factor is recent progress in digital techniques, including the use of graphics processing units and the subsequent acceleration in analysis applications.

## 4. Particular Role/Strengths of Research Groups in Germany

The strength of German radio astronomy has multiple reasons. One of the main ones is the good tradition in training of young astronomers: at present the graduate schools in the Universities, excellence clusters and particularly the International Max Planck Research School for Astronomy and Astrophysics are fostering excellent PhD theses, apart from the good foundation in the master and bachelor level in several universities. A second, important factor is the use of Effelsberg, which serves as a training platform for new techniques and offers access to a world-class facility to German astronomers. The third and most recent pilar of the German community is the participation in LOFAR. This has enlarged the community notably, also with the participation of traditionally disconnected institutions such as the Forschungszentrum Jülich. German scientists contribute not only to harvest the rich LOFAR science cases, but are also pioneering new modes of using LOFAR, such as with the recently developed FACET calibration pipeline, essential for the LOFAR Surveys, and by establishing single station hard- and software to allow the observation of radio pulsars at high cadence. Several groups are intensive users of International facilities such as the JVLA, or the EVN and the Global VLBI arrays. For instance, for the last two semesters of the VLA, a total of 441 proposals were submitted, out of which 308 were approved, and for principal investigators ffiliated in Germany, 31 proposals were submitted and 24 were approved (M. Claussen, priv. comm.).

Given the strong technical component of radio astronomy, the cooperation with Universities of Applied Sciences (Fachochschulen) has been very productive, especially in areas as digitisation, radio frequency interference mitigation, etc. The cooperation with strong theory groups both in the analytic and phenomenology and in the numerical aspects is as well essential for the scientific success of German radio astronomers.

A strong technical expertise is present in the community, and there are important strategic decisions such as the expansion of the LOFAR community, the participation in the MeerKAT telescope with a perspective into the SKA, and the presence of individuals as core team members in SKA subprojects. German radio astronomy is also the backbone for pulsar timing projects such as EPTA and LEAP, which will complement present efforts like LIGO in the detection of low-frequency gravitational waves from merging of (supermassive) black holes. The role of the MPIfR in coordinating European projects such as RadioNet rounds the strength of the German community, also in the area of governance.

## 5. Dominant Science Cases

The dominant science case for the next years have been reported in detail by the ERTRC recently. Existing and new radio telescopes are vital for areas such hydrogen at high redshift, especially for discovering and studying the epochs of the cosmic dawn and of reionization, strong gravity and the discovery of nano-Hertz gravitational waves, compact objects, cosmic magnetism and molecular astrophysics. The SKA will signify a major leap particularly addressing atomic hydrogen and strong gravity. The existing facilities are essential to complement the SKA effort, to perform follow-up observations, and especially concerning the frequency coverage and the sky in the Northern Hemisphere not accessible to the southern antennas. The improvements in instrumentation such as frequency agility, increase in bandwidth, sensitivity in total intensity and polarisation, enhancements in the field of view and spectral resolution will boost radio astronomical progress over the next years.

The different areas of research and related facilities in cm and dm wavelengths over the next years are listed in Table 1.

Table 1: Radio Astronomy topics for the 2020s - a summary and related facilities

| Topic | Eff. | VLBI | LOFAR | SKA |
|---|---|---|---|---|
| Epoch of reionisation and cosmic dawn | | | ✔ | ✔ |
| Gravity and general relativity at the extreme regime, gravitational waves, pulsars | ✔ | | ✔ | ✔ |
| Magnetic fields at all scales | ✔ | ✔ | ✔ | ✔ |
| Life in the universe | ✔ | | | ✔ |
| Transient sources | ✔ | | ✔ | ✔ |
| Active Galactic Nuclei, jets | ✔ | ✔ | ✔ | ✔ |
| Solar physics & space weather | ✔ | | ✔ | ✔ |
| Ultra-high energy cosmic rays | | | ✔ | ✔ |
| Extragalactic surveys | ✔ | ✔ | ✔ | ✔ |
| Astrometry & reference frames | | ✔ | | |
| Cosmology | ✔ | | ✔ | ✔ |

### 6. Summary and Conclusion
To extend the present success, it is important to foster the exchange between the Universities, the Max-Planck-Gesellschaft, the Helmholtz-Gemeinschaft, the Leibniz Association, and the Fraunhofer-Gesellschaft. Effelsberg, the flagship of German radio astronomy, will remain at the top in sensitivity and polarisation capabilities from 1 cm to 10 cm wavelengths over the next decade, and serves with its continuous enhancements as a test bed for new technology. LOFAR is preparing to become LOFAR 2.0 in order to remain the world leading facility at low radio frequencies, at least until SKA-LOW becomes fully operational by the mid 2020s. It will then remain complementary in covering the Northern Hemisphere. A careful assessment of the funding schemes for large project such as the SKA is needed, and the connection with the information technology community, concerning software development in the areas of big data and data analytics needs to be intensified. Further development of cooperation and governance practices will be needed in the moving landscape towards large facilities. Synergies with other facilities, as described e.g., in the high-energy contribution, are significant. In sum, German radio astronomy in cm wavelengths is in good shape, it makes outstanding science with an excellent use of Effelsberg, LOFAR, networks for VLBI and pulsar timing, and the focus will be put into SKA science for the next years.

### Acknowledgments
We acknowledge useful comments and input from a large number of individuals including R. Beck, H. Beuther, P. Biermann, M. Brüggen, A. Brunthaler, M. Kramer, A. Kraus, M. Krause, T.P. Krichbaum, A.P. Lobanov, A. Mao, R.W. Porcas, N. Wex, O. Wucknitz, and J. Anton Zensus.

---

[i] See http://www.astronet-eu.org/FP6/astronet/www.astronet-eu.org/spip33fa.html?rubrique27
[ii] See http://ertrc.strw.leidenuniv.nl/
[iii] Members of the GLOW Consortium are the Universities of Bielefeld, Bochum, Bonn, Cologne, Hamburg and Würzburg, Landessternwarte Thüringen (Tautenburg), Max-Planck-Institutes in Bonn (MPIfR) and Garching (MPA), the AIP Potsdam, the FZ Jülich and the Excellence Cluster Universe (LMU and TU Munich).

[iv] See Antoniadis et al., *A Massive Pulsar in a Compact Relativistic Binary*, Science 340, 448 (2013), http://dx.doi.org/10.1126/science.1233232

[v] See Reid et al., *Trigonometric Parallaxes of Massive Star-Forming Regions. VI. Galactic Structure, Fundamental Parameters, and Noncircular Motions*, The Astrophysical Journal 700, 137-148, 2009, http://dx.doi.org/10.1088/0004-637X/700/1/137, and Brunthaler et al., *The Bar and Spiral Structure Legacy (BeSSeL) survey: Mapping the Milky Way with VLBI astrometry*, Astronomische Nachrichten, 332, 461, 2011, http://dx.doi.org/10.1002/asna.201111560

[vi] See Walter et al., *THINGS: The H I Nearby Galaxy Survey*, The Astronomical Journal, 136, 2563-2647, 2008, http://dx.doi.org/10.1088/0004-6256/136/6/2563

[vii] See http://www.physics.purdue.edu/astro/MOJAVE/ and Zamaninasab et al., *Dynamically important magnetic fields near accreting supermassive black holes*, Nature, 510, 126-128, 2014, http://dx.doi.org/10.1038/nature13399; Savolainen et al., *Relativistic beaming and gamma-ray brightness of blazars*, Astronomy & Astrophysics, 512, A24, 2010, http://dx.doi.org/10.1051/0004-6361/200913740; Lister et al., *MOJAVE: Monitoring of Jets in Active Galactic Nuclei with VLBA Experiments. V. Multi-Epoch VLBA Images*, Astronomical Journal, 137, 3718-3729, 2009, http://dx.doi.org/10.1088/0004-6256/137/3/3718 ; Lister et al., *MOJAVE: Monitoring of Jets in Active Galactic Nuclei with VLBA Experiments. VI. Kinematics Analysis of a Complete Sample of Blazar Jets*, Astronomical Journal, 138, 1874-1892, 2009, http://dx.doi.org/10.1088/0004-6256/138/6/1874; Lister et al., *A Connection Between Apparent VLBA Jet Speeds and Initial Active Galactic Nucleus Detections Made by the Fermi Gamma-Ray Observatory*, Astrophysical Journal 696, L22-L26, 2009, http://dx.doi.org/10.1088/0004-637X/696/1/L22; Lister et al., *MOJAVE. X. Parsec-scale Jet Orientation Variations and Superluminal Motion in Active Galactic Nuclei*, Astronomical Journal, 146, 120, 2013, http://dx.doi.org/10.1088/0004-6256/146/5/120; Lister et al., *MOJAVE XIII. Parsec-Scale AGN Jet Kinematics Analysis Based on 19 years of VLBA Observations at 15 GHz*, Astronomical Journal, in press, 2016, http://arxiv.org/abs/1603.03882 .

[viii] See Eatough et al., *A strong magnetic field around the supermassive black hole at the centre of the Galaxy*, Nature 501, 391-394, 2013, http://dx.doi.org/10.1038/nature12499, press release in https://www.mpg.de/7502957/magnetar_milkyway

[ix] See http://www3.mpifr-bonn.mpg.de/div/vlbi/fgamma/fgamma.html and Fuhrmann et al., *Detection of significant cm to sub-mm band radio and γ-ray correlated variability in* Fermi *bright blazars*, Monthly Notices of the Royal Astronomical Society, 441, 1899-1909, 2014, http://dx.doi.org/10.1093/mnras/stu540, press release under http://www.mpg.de/1218670/Aktive_Galaxienkerne?c=166022

[x] See https://astro.uni-bonn.de/~jkerp/index.php?page=EBHISproject and Winkel et al., *The Effelsberg-Bonn HI Survey: Milky Way gas. First data release*, Astronomy & Astrophysics, 585, A41, 2016, http://dx.doi.org/10.1051/0004-6361/201527007

[xi] See Kramer & Champion, *The European Pulsar Timing Array and the Large European Array for Pulsars*, Classical and Quantum Gravity 30, 224009, 2013, http://adsabs.harvard.edu/abs/2013CQGra..30v4009K

[xii] See Beck, *Magnetism in the spiral galaxy NGC 6946: magnetic arms, depolarization rings, dynamo modes, and helical fields*, Astronomy & Astrophysics 470, 539-556, 2007, http://dx.doi.org/10.1051/0004-6361:20066988

[xiii] See Keane et al., *The host galaxy of a fast radio burst*, Nature 530, 453-456, 2016, http://dx.doi.org/10.1038/nature17140

[xiv] See Buitink et al., *A large light-mass component of cosmic rays at $10^{17}$–$10^{17.5}$ electronvolts from radio observations,* Nature 531, 70-73, 2016, http://dx.doi.org/10.1038/nature16976

[xv] See Rubart & Schwarz, *Cosmic radio dipole from NVSS and WENSS, Astronomy & Astrophysics* 555, A17, 2013, http://dx.doi.org/10.1051/0004-6361/201321215

[xvi] See Fletcher et al., *Magnetic fields and spiral arms in the galaxy M 51*, MNRAS 412, 2396, 2011, http://dx.doi.org/10.1111/j.1365-2966.2010.18065.x

[xvii] See de Gasperin et al., *M 87 at metre wavelengths: the LOFAR picture*, Astronomy & Astrophysics 547, A56, http://dx.doi.org/10.1051/0004-6361/201220209

[xviii] See Breitling et al., *The LOFAR Solar Imaging Pipeline and the LOFAR Solar Data Center*, Astronomy & Computing 13, 99, 2015, http://dx.doi.org/10.1016/j.ascom.2015.08.001